\documentclass[conference]{IEEEtran} 

\usepackage{cite}
\usepackage{amsmath,amssymb,amsfonts}
\usepackage{algorithmic}
\usepackage{graphicx}
\usepackage{textcomp}
\usepackage{xcolor}
\usepackage{mathtools}
\def\BibTeX{{\rm B\kern-.05em{\sc i\kern-.025em b}\kern-.08em
    T\kern-.1667em\lower.7ex\hbox{E}\kern-.125emX}}

\usepackage{algorithm}
\usepackage{algorithmic}

\usepackage{color}
\usepackage{tablefootnote}

\usepackage{amsmath,amssymb,amsthm}
\usepackage{multirow}

\newcommand{\mbf}{\mathbf}

\newcommand{\trace}{\operatorname{trace}}

\usepackage{algorithm}
\usepackage{algorithmic}
\usepackage{longtable,tabularx}

\begin{document}

\title{Propagation of Uncertainty with the Koopman Operator}

\author{

\IEEEauthorblockN{Simone Servadio,}
\IEEEauthorblockA{\textit{Iowa State University}, Ames, IA, USA}
\thanks{Dr. Simone Servadio, Assistant Professor, Department of Aerospace Engineering, servadio@iastate.edu}

\IEEEauthorblockN{Giovanni Lavezzi,}
\IEEEauthorblockA{\textit{Massachusetts Institute of Technology}, Cambridge, MA, USA}
\thanks{Dr. Giovanni Lavezzi, Postdoctoral Associate, Department of Aeronautics and Astronautics, glavezzi@mit.edu}

\IEEEauthorblockN{Christian Hofmann,}
\IEEEauthorblockA{\textit{Massachusetts Institute of Technology}, Cambridge, MA, USA}
\thanks{Dr. Christian Hofmann, Postdoctoral Associate, Department of Aeronautics and Astronautics, hofmannc@mit.edu}

\IEEEauthorblockN{Di Wu,}
\IEEEauthorblockA{\textit{Massachusetts Institute of Technology}, Cambridge, MA, USA}
\thanks{Dr. Di Wu, Postdoctoral Associate, Dept of Aeronautics and Astronautics, d\_wu@mit.edu}

\IEEEauthorblockN{Richard Linares,}
\IEEEauthorblockA{\textit{Massachusetts Institute of Technology}, Cambridge, MA, USA}
\thanks{Dr. Richard Linares, Rockwell International Career Development Associate Professor, Dept. of Aeronautics and Astronautics, linaresr@mit.edu}
}

\markboth{Transactions on Aerospace and Electronic Systems}%
{Shell \MakeLowercase{\textit{et al.}}: Bare Demo of IEEEtran.cls for IEEE Journals}

\maketitle
\thispagestyle{plain}
\pagestyle{plain}

\begin{abstract}
This paper proposes a new method to propagate uncertainties undergoing nonlinear dynamics using the Koopman Operator (KO). Probability density functions are propagated directly using the  Koopman approximation of the solution flow of the system, where the dynamics have been projected on a well-defined set of basis functions. The prediction technique is derived following both the analytical (Galerkin) and numerical (EDMD) derivation of the KO, and a least square reduction algorithm assures the recursivity of the proposed methodology. 
\end{abstract}


\IEEEpeerreviewmaketitle


\section{Introduction}
\IEEEPARstart{T}{he} accurate propagation of uncertainty in stochastic systems is a problem that still attracts attention since a closed-form solution is not available whenever the dynamics are nonlinear. In the case of a linear and Gaussian system, the Gaussian distributions maintain their Gaussianity, and the propagated mean and covariance of the probability density function (PDF) are calculated exactly, following the Kalman filter prediction step \cite{kal}. However, most processes are nonlinear, and the initial spread of uncertainty is doomed to increase as time continues \cite{ss}. The first and easiest solution is to linearize the system at the most current estimate and propagate the PDF as if the system were linear, evaluating Jacobians \cite{gelb}. Although working, the accuracy of this solution becomes extremely poor whenever nonlinearities are strong \cite{singla}. Therefore, more precise techniques have been developed to address this problem and to provide a more reliable prediction of the propagated distribution. The unscented transformation (UT) relies on the Gaussian assumption of the initial PDF \cite{ut}. It selects well-chosen sigma points based on the covariance, and the predicted PDF is reconstructed from the propagated points. The UT has shown higher accuracy results when compared to basic linearization \cite{nonl}. Another approach consists of the use of State Transition Tensors (STT) \cite{park} that fully rely on the nonlinear mapping of mean and covariance to achieve a more accurate prediction than linearization. A similar approach has been obtained using Differential Algebra (DA) techniques \cite{bertzda}, where the tensor mathematics has been replaced by Taylor polynomials, eliminating the need to evaluate the high-order tensors directly \cite{dap}. Another solution is the Polynomial Chaos Expansion (PCE) method, which is often used when there is a requirement for accurate determination of higher statistical moments or the complete PDF. This approach involves approximating input parameters and their corresponding solutions through a series expansion that relies on orthogonal polynomials \cite{PCE}. Despite its suitability for non-Gaussian models, the PCE method suffers from the curse of dimensionality. 

This work presents a new technique to propagate uncertainties for Hamiltonian nonlinear systems \cite{ham} thanks to the Koopman Operator (KO). Developed by Koopman \cite{ko1,ko2}, the operator represents a finite nonlinear system as a linear infinite dimensional system, where the dynamics have been projected on a set of eigenfunctions. In this paper, the time propagation of the nonlinear dynamics is operated directly into the PDF using the Koopman approximation of the solution flow. Thus, the new technique is analytical and integrates the full PDF function rather than its approximation via central moments, providing the full shape of the distribution. 

\section{Propagation of Uncertainty in Hamiltonian Systems} \label{sec1}
Assume a Hamiltonian system governed by a set of nonlinear ordinary differential equations ${\bf f} (\mbf x)$, where $\mbf x \in \mathbb R^n$ is the state that depends on the time evolution $t$, and ${\bf f }:\mathbb R^n\rightarrow \mathbb R^n$ is the model of the nonlinear dynamics, with $n$ the number of dimensions in which the problem is defined. Given an initial condition, $\mbf x_0$, of the system at time $t_0$, the initial value problem is written in the form:
\begin{equation}
    \left\{ \begin{tabular}{l}
    $\displaystyle\frac{\partial}{\partial t}{\bf x}(t) = {\bf f}(t,{\bf x})$ \\
    ${\bf x}(t_0) = {\bf x_0}$ 
\end{tabular}  \right.
\end{equation}
The solution flow of the system that propagates the state from $t_0$ to a generic time $t$ can be expressed as 
\begin{equation}
\mbf x(t) = \mathcal F(\mbf x_0; t_0\rightarrow t)
\end{equation}
Therefore, by substitution
\begin{equation}
\frac{d}{dt} \mathcal F(\mbf x_0; t_0\rightarrow t) = {\bf f}(t,\mathcal F(\mbf x_0; t_0\rightarrow t))
\end{equation}
It becomes evident that the solution flow respects the following properties:
\begin{align}
    \mathcal F(\mbf x_0; t_0\rightarrow t_0) &= \mbf x_0 \\
    \mathcal F(\mathcal F(\mbf x_0; t_0\rightarrow t); t\rightarrow t_0) &= \mbf x_0 \label{eq:1}
\end{align}
where the latter equation merely states that the functional inverse is obtained by inverting the time arguments of the flow, analogously on how states are propagated forward and backward in time using the state transition matrix and its inverse. Thus, calling with $\mbf x_f$ the state at the final time $t_f$, the following identities hold:
\begin{align}
    \mbf x_f &= \mathcal F(\mbf x_0; t_0\rightarrow t_f) \\
    \mbf x_0 &= \mathcal F(\mbf x_f; t_f\rightarrow t_0) 
\end{align}
By differentiating Eq. \eqref{eq:1} with respect to time, it is possible to evaluate the rate of change of the solution flow.
\begin{align}
    \dfrac{d}{dt} \mathcal F(& \mathcal F(\mbf x_0; t_0\rightarrow t); t\rightarrow t_0) = \mbf 0  \\
    =& \dfrac{\partial}{\partial t} \mathcal F(\mathcal F(\mbf x_0; t_0\rightarrow t); t\rightarrow t_0) +\nonumber \\
    &\ \ \ \ \ \dfrac{\partial}{\partial \boldsymbol \chi}  \mathcal F(\boldsymbol \chi; t\rightarrow t_0) \dfrac{\partial}{\partial t}\mathcal F(\mbf x; t_0\rightarrow t)
\end{align}
which can be rewritten by substituting the dynamics  
\begin{equation}
    \dfrac{\partial}{\partial t} \mathcal F(\mbf x; t\rightarrow t_0) = - \dfrac{\partial}{\partial \boldsymbol \chi} \mathcal F(\boldsymbol \chi; t\rightarrow t_0) \bigg|_{\boldsymbol \chi = \mbf x} {\bf f}(t,{\bf x}) \label{eq:2}
\end{equation}
Most systems are not deterministic, and it is crucial to consider the stochastic nature of the dynamics. Therefore, taking the initial condition of the state as a random vector with known PDF $p_{\mbf x_0}(\mbf x_0,t_0)$, the distribution evolves according to the Fokker-Plank-Kolmogorov (FPK) \cite{fpk} equation without diffusion:
\begin{align}
    &\dfrac{\partial}{\partial t} (p_{\mbf x(t)}(\mbf x,t)) = - \sum^n_{i=1} \dfrac{\partial}{\partial x_i}\left( f_i(t,\mbf x) p_{\mbf x(t)}(\mbf x,t)\right) \nonumber \\
    =& - \sum^n_{i=1} f_i(t,\mbf x) \dfrac{\partial ( p_{\mbf x(t)}(\mbf x,t))}{\partial x_i} - \sum^n_{i=1} p_{\mbf x(t)}(\mbf x,t) \dfrac{\partial ( f_i(t,\mbf x)) }{\partial x_i} \nonumber \\
    &= - \dfrac{\partial ( p_{\mbf x(t)}(\mbf x,t))}{\partial \mbf x} \mbf f (t,\mbf x) - p_{\mbf x(t)}(\mbf x,t)\trace \left( \dfrac{\partial \mbf f (t,\mbf x)}{\partial \mbf x} \right)
\end{align}
This formulation, also known as the Louville equation, is derived after applying the chain rule and vectorizing the formulation. In particular, for Hamiltonian systems, the trace of the Jacobian of $\mbf f(t,\mbf x)$ is null, leading to
\begin{align}
    \dfrac{\partial}{\partial t} (p_{\mbf x(t)}(\mbf x,t)) + \dfrac{\partial ( p_{\mbf x(t)}(\mbf x,t))}{\partial \mbf x} \mbf f (t,\mbf x) = \dfrac{d}{dt} (p_{\mbf x(t)}(\mbf x,t)) = \mbf 0
\end{align}
Therefore, it is possible to show that 
\begin{align}
     &\dfrac{d}{dt} (p_{\mbf x(t)}(\mbf x,t)) =  \dfrac{d}{dt} \left(p_{\mbf x_0}(\mathcal F(\mbf x; t\rightarrow t_0),t)\right) \\
     &= \dfrac{\partial p_{\mbf x_0}(\boldsymbol \chi,t)}{\partial \boldsymbol \chi}\left(  \dfrac{\partial \mathcal F(\mbf x; t\rightarrow t_0) }{\partial t}  +  \dfrac{\partial \mathcal F(\mbf x; t\rightarrow t_0)}{\partial \mbf x}  {\bf f}(t,{\bf x})\right)
\end{align}
where the terms in parenthesis add to zeros due to Eq. \eqref{eq:2}, proving that for any time 
\begin{align}
    p_{\mbf x(t)}(\mbf x,t) &= p_{\mbf x(t_0)}(\mathcal F(\mbf x; t\rightarrow t_0),t_0) \nonumber \\
    &= p_{\mbf x(t_0)}((\mathcal F(\mbf x; t_0\rightarrow t))^{-1},t_0)  \label{eq:3}
\end{align}
where $(\mathcal F(\mbf x; t_0\rightarrow t))^{-1} =  \mathcal F(\mbf x; t\rightarrow t_0) $ indicates the inverted solution flow that goes back from the given time to the initial one. That is, Eq. \eqref{eq:3} indicates that the state PDF is propagated forward in time by evaluating the initial PDF with the inverted solution flow \cite{damap}. 

The Koopman Operator offers a technique to represent and approximate the exact solution flow as a linear combination of the eigenfunctions of the systems. This approximation has been used, so far in the literature, merely to propagate the state of the system and extract observables. However, when analyzed, the KO solution flow can be easily inverted and, therefore, used to propagate uncertainties for Hamiltonian systems.

\section{Generating the Koopman Solution Flow} \label{sec2}
The KO provides an approximation of the solution flow of the dynamics by projecting them on a set of well-defined basis functions. The main idea is to obtain the time behavior of such basis functions such that the propagation of any observable of the state can be expressed as a linear combination of such functions. Two main approaches have been derived in the literature to obtain the Koopman eigendecomposition of the dynamics: the analytical approach via the Galerkin method \cite{servaKO1,servaKO2,servaKO3,servaKO4,servaKO5,servaKO6,servaKO7} and the numerical approach using Extended Dynamic Mode Decomposition (EDMD) \cite{edmd1,edmd2,edmd3,edmd4}. A quick overview is proposed here without going too deep in detail, as theory has been covered exhaustively in recent publications. 

\subsection{The Analytical Approach}
Given the dynamics of an autonomous system $\mbf f(\mbf x)$, the Koopman Operator $\mathcal K(\cdot) $ is an infinite dimensional operator that describes how an observable function of the state $\mbf g (\mbf x) \in \mathcal G(\mbf x)$, being $\mathcal G(\mbf x)$ all observable functions, evolves in time
\begin{equation}
\mathcal{K}\left(\mbf g({\bf x})\right) = \frac{d}{dt}\mbf g({\bf x}) = \left( \nabla_{{\bf x}} \mbf g({\bf x})\right)\frac{d}{dt}{\bf x}(t) = \left( \nabla_{{\bf x}} \mbf g({\bf x})\right){\bf f}({\bf x})
\end{equation}
which can be seen as the application of the chain rule. If we were to describe the observable as a linear combination of basis functions, which time behavior under the system's dynamics is known, then the state propagation of the system would be solved by selecting the identity observable, extracting the state \cite{servaKO4}. Therefore, after selecting a set of $\eta$ basis functions, in this case the Legendre polynomials $\mathcal L(\mbf x)$, the goal of the approach is to derive the Koopman matrix $\mbf K$ that describes linearly how the rate of change of the basis function is dependent on the functions themselves:
\begin{equation}
    \dfrac{d\mathcal L(\mbf x)}{d t} = {\mbf K} \mathcal L(\mbf x) \label{eq:ori}
\end{equation}
which is a system of $\eta$ equations. To evaluate the entry of each component of the Koopman matrix, we must first analyze how the $i$-th ODE is written \cite{servaKO6}:
\begin{align}
\dfrac{d \mathcal L_i(\mbf x)}{d t} &= \bigg\langle \dfrac{d \mathcal L_i(\mbf x)}{d t} , \mathcal L_0(\mbf x)\bigg\rangle \mathcal L_0(\mbf x)  \nonumber \\  
& \quad \quad  + \bigg\langle \dfrac{d \mathcal L_i(\mbf x)}{d t} , \mathcal L_1(\mbf x)\bigg\rangle  \mathcal L_1(\mbf x) +    \dots \nonumber\\
&= \sum_{j=0}^{\eta} \bigg\langle \dfrac{d \mathcal L_i(\mbf x)}{d t} , \mathcal L_j(\mbf x)\bigg\rangle \mathcal L_j (\mbf x)
\end{align} 
where each inner product $\langle \cdot , \cdot \rangle$ is the entry $\mbf{ K}_{ij}$  
\begin{equation}
\mbf{ K}_{ij}=\bigg \langle \dfrac{d \mathcal L_i(\mbf x)}{d t} , \mathcal L_j(\mbf x)\bigg \rangle
\end{equation}
that expresses the projection of the derivative of the basis function onto the orthonormal set of basis functions. The total derivative of the basis functions with respect to time can be expanded to show the influence of the dynamics
\begin{align}
\dfrac{d \mathcal L_i(\mbf x)}{d t} &= \dfrac{\partial \mathcal L_i(\mbf x)}{\partial x_1} \dfrac{d x_1}{d t} + \dfrac{\partial \mathcal L_i(\mbf x)}{\partial x_2} \dfrac{d x_2}{d t}  +  \dots  \nonumber\\
&= \sum_{j=1}^{n} \dfrac{\partial \mathcal L_i(\mbf x)}{\partial x_j} \dfrac{d x_j}{d t} \nonumber\\
&= \sum_{j=1}^{n} \dfrac{\partial \mathcal L_i(\mbf x)}{\partial x_j}{ f}_j({\bf x}). 
\end{align} 
where the terms $\dfrac{\partial \mathcal L_i(\mbf x)}{\partial x_j}$ are easy to compute since they are polynomial derivatives. Therefore, the Koopman matrix is evaluated component-wise by calculating the inner products using the Galerkin method, where 
\begin{equation}
\bigg \langle   \dfrac{d \mathcal L_i(\mbf x)}{d t}, \mathcal L_j(\mbf x)\bigg \rangle =\int_{\mathcal X}  \dfrac{d \mathcal L_i(\mbf x)}{d t}\mathcal L_j(\mbf x) w(\mbf x)d \mbf x 
\end{equation}
with $w(\mbf x)$ a positive weighting function defined on the space domain $\mathcal X \subseteq \mathbb R^n$, selected as the uniform distribution for the results proposed in this paper \cite{servaKO2,servaKO1}. The Koopman matrix tells how the system changes in time as a linear combination of the basis functions. From this information, it is possible to extract the time history of any given observable of the state $\mbf g (\mbf x) $ by following the same approach of describing the observable as a combination of the basis function
\begin{equation}
    \mbf g(\mbf x) = \sum_{j=1}^{\eta} \langle \mbf g_i (\mbf x), \mathcal L_j(\mbf x)\rangle  \mathcal L_j(\mbf x) = \mbf H \mathcal L(\mbf x) \label{merge1}
\end{equation}
This is achieved, once again, by the use of the Galerkin method by evaluating the inner products
\begin{equation}
    \mbf H_{i,j} = \langle \mbf g_i (\mbf x), \mathcal L_j(\mbf x)\rangle
\end{equation}
that define the observable matrix $\mbf H$, with dimensions $\gamma \times \eta$, where $\gamma$ is the number of observables. The state of the system is extracted by selecting the identity observable \cite{servaKO4,servaKO6}. 

The Koopman approximation of the solution flow is derived by performing the eigendecomposition of $\mbf K$:
\begin{equation}
    \mbf V \mbf K = \boldsymbol{\Lambda} \mbf V \label{eq:eig}
\end{equation}
where $\boldsymbol{\Lambda}$ is the matrix of eigenvalues and the left eigenvectors matrix $\mbf V$ defines the eigenfunctions of the system 
\begin{equation}
    \boldsymbol \phi (\mbf x) = \mbf V \mathcal L (\mbf x) \label{merge2}
\end{equation} 
The dynamics are diagonal with respect to the eigenfunctions, meaning that the propagation is easy to compute. Given initial and final time, $t_0$ and $t_f$ respectively, the propagated eigenfunctions are
\begin{equation}
    \boldsymbol \phi (\mbf x(t_f)) =  \exp ( \Delta t \boldsymbol \Lambda) \boldsymbol \phi (\mbf x(t_0))  \label{merge3}
\end{equation}
with $\Delta t = t_f-t_0$, since the system has been diagonalized
\begin{align}
    \dfrac{d}{dt}&\boldsymbol \phi (\mbf x(t))  = \dfrac{d}{dt}\mbf V \mathcal L (\mbf x(t)) = \mbf V \dfrac{d}{dt}\mathcal L (\mbf x(t))  \nonumber\\
    &= \mbf V \mbf K \mathcal L (\mbf x(t)) = \boldsymbol{\Lambda} \mbf V  \mathcal L (\mbf x(t)) = \boldsymbol{\Lambda} \boldsymbol \phi (\mbf x(t))
\end{align}
having used Eq. \eqref{eq:ori} and Eq. \eqref{eq:eig} for the last two substitutions. By merging Eq. \eqref{merge1} with Eq. \eqref{merge2}, its inverse, and Eq. \eqref{merge3}, the final KO solution is derived
\begin{align}
    \mbf g(\mbf x_f) &= \mbf H \mathcal L (\mbf x_f) \nonumber\\
    &= \mbf H \mbf V^{-1} \boldsymbol{\phi }(\mbf x_f) \nonumber\\
    &= \mbf H \mbf V^{-1} \exp(\Delta t \boldsymbol{\Lambda})\boldsymbol{\phi }(\mbf x_0) \nonumber\\
    &= \mbf H \mbf V^{-1} \exp(\Delta t \boldsymbol{\Lambda})\mbf V \mathcal L(\mbf x_0) \label{prop}
\end{align}
This last formulation shows, step-by-step, every passage of the Koopman approximation \cite{servaKO7}. First, the basis functions are evaluated at the given state $\mbf x_0 = \mbf x(t_0)$; second, the eigenfunctions at the initial time are evaluated by the eigenvectors rotation $\mbf V$; third, eigenfunctions are propagated to $t_f$ using the exponential; fourth, the inverse rotation $\mbf V^{-1}$ brings the basis functions at the final state $\mbf x_f = \mbf x(t_f)$; and lastly, matrix $\mbf H$, extracts the observable as linear combination of the basis functions \cite{servaKO6}. This formulation takes the name of the Koopman Operator State Transition Polynomial Map (KOSTPM). The full process is derived in detail in the provided references.

\subsection{The Numerical Approach}
Due to its wide use in literature, the EDMD algorithm has been selected as the numerical method to approximate the KO and derive the discrete Koopman matrix. The EDMD calculates the Koopman matrix by fitting the operator to training data. Therefore, consider a data set of snapshots pairs, i.e., $\{(\mbf x_m,\mbf y_m)\}_{m=1}^M$, that are organized into two separate data sets:
\begin{align}
    \mbf X = \begin{bmatrix}
        \mbf x_1 & \mbf x_2 & \dots & \mbf x_M
    \end{bmatrix} \\
    \mbf Y = \begin{bmatrix}
        \mbf y_1 & \mbf y_2 & \dots & \mbf y_M
    \end{bmatrix} 
\end{align}
where both $\mbf x_m \in \mathcal X$ and $\mbf y_m \in \mathcal X$ are snapshots of the state of the system, with $\mbf y_m = \mathcal F (\mbf x_m,t_0\rightarrow t)$. After selecting a set of basis functions, in the presented case, the Legendre polynomials $\mathcal L(\mbf x)$, the EDMD algorithm uses a least square approach to fit the propagated functions to their Koopman approximation. Therefore, after evaluating the basis functions at the given snapshots
\begin{align}
    \mathcal L(\mbf X) = \begin{bmatrix}
         \mathcal L(\mbf x_1) &  \mathcal L(\mbf x_2) & \dots &  \mathcal L(\mbf x_M)
    \end{bmatrix} \\
     \mathcal L(\mbf Y) = \begin{bmatrix}
         \mathcal L(\mbf y_1) &  \mathcal L(\mbf y_2) & \dots &  \mathcal L(\mbf y_M)
    \end{bmatrix} 
\end{align}
where 
\begin{equation}
    \mathcal L(\mbf x_1) = \begin{bmatrix}
         \mathcal L_1(\mbf x_1) &  \mathcal L_2(\mbf x_1) & \dots &  \mathcal L_\eta(\mbf x_1)
    \end{bmatrix}^T
\end{equation}
with $\eta$ the number of Legendre polynomials, the KO is evaluated by solving the least square problem \cite{edmd1}
\begin{align}
    \min_{\tilde{\mbf K} \in \mathbf R^{\eta \times \eta}} \mathcal J_{KO} =& \min_{\tilde{\mbf K} \in \mathbf R^{\eta \times \eta}} \vert \vert  \mathcal L(\mbf Y) - \tilde{\mbf K} \mathcal L(\mbf X)\vert \vert ^2_F \nonumber\\
    =& \min_{\tilde{\mbf K} \in \mathbf R^{\eta \times \eta}} \sum_{i=1}^M \vert \vert  \mathcal L(\mbf y_i)- \tilde{\mbf K} \mathcal L(\mbf x_i) \vert \vert ^2_F 
\end{align}
which solution is
\begin{equation}
    \tilde{\mbf K} = (\mbf Q^T \mbf Q)^{-1}\mbf Q^T \mbf P
\end{equation}
where
\begin{align}
    \mbf Q = \dfrac{1}{M}\sum_{m=1}^M \mathcal L(\mbf x_m)^T \mathcal L(\mbf x_m) \label{eq:10} \\
    \mbf P = \dfrac{1}{M}\sum_{m=1}^M \mathcal L(\mbf x_m)^T \mathcal L(\mbf y_m) \label{eq:11}
\end{align}
As a result, matrix $\tilde{\mbf K}$ is the discrete-time approximation of the KO trained at given snapshots. In a similar manner, the observable matrix $\tilde{\mbf H}$ is obtained by inverting the relationship in Eq. \eqref{merge1}
\begin{equation}
    \tilde{\mbf H} =(\mathcal L(\mbf X)^T \mathcal L(\mbf X))^{-1} \mathcal L(\mbf X)^T \mbf g(\mbf X) 
\end{equation}
In this way, the eigendecomposition of the Koopman matrix $\tilde{\mbf K}$ retrieved by EDMD can be performed as per Eq. \eqref{eq:eig}, and, given the observable matrix $\tilde{\mbf H}$, the final solution can be computed using Eq. \eqref{prop}. In this paper, a modified version of the EDMD algorithm (including the selected basis functions) provided as a Matlab library in \cite{mdpi} is used, where the authors proposed two different reduction methods to improve the numerical stability and reduce the number of observables in the EDMD algorithm.

\subsection{The Convergence of the Approaches}
This paper proposes two separate approaches to underline that they hold the same result and that the PDF propagation technique is achieved regardless of the approach. In fact, it is proven that, for a large amount of training data, the EDMD converges to the Galerkin method. This can be seen by looking at Eq. \eqref{eq:10} and Eq. \eqref{eq:11} as the numerical approximation of two integrals
\begin{align}
    \mbf Q_{i,j} &= \int_{\mathcal X} \mathcal L_i(\mbf x) \mathcal L_j(\mbf x) w(\mbf x) d\mbf x =  \langle \mathcal L_i(\mbf x), \mathcal L_j(\mbf x)\rangle\\
    \mbf P_{i,j} &= \int_{\mathcal X} \mathcal L_i(\mbf x) \mathcal L_j(\mathcal F(\mbf x)) w(\mbf x) d\mbf x = \langle  \mathcal L_i(\mbf x), \mathcal K \mathcal L_j(\mbf x) \rangle
\end{align}
The integral accuracy increases as the dataset enlarges to infinity, meaning that the EDMD converges to the Galerkin approach \cite{edmd1}. After all, the Riemann definition of the integral operator is the limit of a summation \cite{reimann}.

The discrete Koopman matrix $\mbf{\tilde K}$ expresses the time evolution of the basis functions for a specific time-step $\Delta t$, at which snapshots are taken. Its relation to the correspondent continuous time Koopman matrix $\mbf K$ derived by the Galerkin approach is
\begin{equation}
    \mbf{\tilde K} = \mbf V^{-1} \exp({\mbf \Lambda \Delta t}) \mbf V
\end{equation}
which, using properties of the matrix exponential $\mbf A \exp({\mbf B}) \mbf A^{-1} = \exp(\mbf A \mbf B \mbf A^{-1})$, holds
\begin{equation}
    \log \mbf{\tilde K} = \log \Big(  \exp(\mbf V^{-1}\mbf \Lambda \Delta t\mbf V)  \Big) = \mbf V^{-1}\mbf \Lambda \Delta t\mbf V = \mbf K \Delta t
\end{equation}
leading to the following relationship to switch from the discrete-time to the continuous-time Koopman matrix
\begin{equation}
    \mbf K = \dfrac{\log \mbf{\tilde K}}{\Delta t}  \label{eq:9}
\end{equation}
It is obvious that the two formulations, EDMD and Galerkin, represent the same structure of the solution 
\begin{align}
    \mbf x_{k+1}  = \mbf H \mbf V^{-1} \exp (\Delta t \boldsymbol \Lambda) \mbf V  \mathcal L (\mbf x_k) = \mbf H \mbf{\tilde K}   \mathcal L (\mbf x_k)
\end{align}
In conclusion, either approach can be selected to derive the novel derivation of the propagation of uncertainty.

\section{The KO map inversion} \label{sec3}
Regardless of the selected technique to obtain the KOSTPM and relative quantities and variables, both the EDMD and the Galerkin formulation can be written by highlighting the dependence of the flow on time and on the initial condition. Indeed, given a desired integration time, $\Delta t = t_f - t_0$, and the state initial condition at $t_0$, $\mbf x_0$, the propagated state at the final time $t_f$, $\mbf x_f$, is evaluated as
\begin{equation}
    \mbf x_f = \mathcal M_{t_0\rightarrow t_f}(\mbf x_0) = \mbf H \mbf V^{-1} \exp (\Delta t \boldsymbol \Lambda) \mbf V  \mathcal L (\mbf x_0)   
\end{equation}
The $\mathcal M_{t_0\rightarrow t_f}(\mbf x_0)$ map is made of orthonormal polynomials, and it can be seen as the KO equivalent of the STT \cite{stt} that connects final to initial state. In order to propagate PDFs according to Eq. \eqref{eq:3}, the map needs to be inverted, such that it provides the initial state given the final one, propagating backward in time. While commonly hard to derive in the general polynomial case \cite{inverse}, the polynomial map inversion becomes trivial with the Koopman mathematics. Thus, thanks to the eigendecomposition of the system and the explicit dependency of the KOSTPM on time, it is sufficient to switch the time boundaries to achieve backward integration:
\begin{align}
 \mathcal W_{t_f\rightarrow t_0} (\mbf x_f) =& \big( \mathcal M_{t_0\rightarrow t_f}(\mbf x_0) \big)^{-1} \nonumber \\
 =& \mbf H \mbf V^{-1} \exp (-\Delta t \boldsymbol \Lambda) \mbf V  \mathcal L (\mbf x_f)   \label{eq:4}
\end{align}
where $\mathcal W_{t_f\rightarrow t_0}(\mbf x_f)$ indicates the inverted map. Therefore, once the dynamics have been projected onto the set of basis functions and analyzed, the evaluation of the retrograde solution flow becomes surprisingly easy. 

\section{The Koopman Propagation of PDFs}\label{sec4}
The Koopman representation of the dynamics can now be applied to propagate distributions. During the mathematical derivation of the theory reported so far, it appears evident that two separate roads can be followed when dealing with a PDF. On one side, the distribution can be seen as an additional observable to be evaluated, while on the other side, the PDF can be propagated using the properties of Hamiltonian systems in the FPK equation. Both cases are here derived and reported.

\subsection{The PDF propagation as an observable} 
Assume an initial distribution for the state of the system, $\psi_{\mbf x}(\mbf x,t_0)$. The PDF is expressed as a function of the state $\psi_{\mbf x}(\mbf x,t_0):\mathbb R^n\rightarrow \mathbb R$ that can be interpreted as the desired observable for the KO. Thus, depending on the approach selected, either numerical or analytical, a new observable matrix can be evaluated that extracts, from the basis functions, the necessary contributions to evaluate the new observable, populating matrix $\mbf H_p$. This matrix is actually a vector of dimensions $1\times \eta$, with $\eta$ being the number of basis functions, since the output of a PDF is a positive scalar. 

In the numerical case, using EDMD, the components of $\mbf H_p$ are evaluated by simulating multiple times the evolution of the PDF, and extracting the observable matrix following the procedure described before. However, in the analytical approach using the Galerkin method, each entry of $\mbf H_p$ is evaluated using the inner product:
\begin{equation}
    \mbf H_{p,j} = \langle \psi_{\mbf x}(\mbf x,t_0), \mathcal L_j(\mbf x) \rangle  \quad \forall \quad j = 1,\dots,ns
\end{equation}
These integrals are computed either analytically, for easy distributions, or numerically using quadrature for more complex $\psi_{\mbf x}(\mbf x,t_0)$. This latter approach guarantees to describe the evolution of the PDF on the whole state domain, while the EDMD solution is reliable only on the state space subset where it has been trained. Regardless of the approach, the propagated PDF to the final time $t_f$ can be expressed as 
\begin{equation}
    \psi_{\mbf x}(\mbf x,t_f) = \mbf H_p \mbf V^{-1} \exp (\Delta t \boldsymbol \Lambda) \mbf V  \mathcal L (\mbf x)   
\end{equation}

\subsection{The PDF Propagation via Map Inversion}
The propagation of the distribution as an observable has the drawback of requesting the evaluation of additional integrals that might be hard to calculate accurately. Moreover, the addition of more complex training data in the EDMD inevitably decreases the performance of the technique. Therefore, since the initial state PDF is supposed to be known, or approximated, with a well-defined function $\psi_{\mbf x}(\mbf x,t_0)$, it is convenient to use the trivial evaluation of the KOSTPM inversion of Eq. \eqref{eq:4} to propagate uncertainties. This expression allows for the propagation of uncertainty according to Eq. \eqref{eq:3}, which, for the KO mathematics, becomes
\begin{equation}
    \psi_{\mbf x}(\mbf x,t_f) = \psi_{\mbf x}(\mathcal W_{t_f\rightarrow t_0} (\mbf x),t_0) \label{eq:5}
\end{equation}
This equation is a simple function evaluation where the variables of the initial PDF are substituted with the inverted KO polynomials. Conceptually, the information regarding the evolution of systems according to the projected dynamics has been integrated into the distribution, which changes its shape accordingly. 

The main advantage of the inverted approach over the observable function approach is its capability of being recursive. Indeed, if a second propagation were to be performed, the new final PDF would be evaluated simply by repeating Eq. \eqref{eq:5} with the updated times without the need to worry about the training set time domain.

\subsection{The Logarithmic Simplification}
Particular attention can be reserved for those distributions whose dependency on the state is in the exponential term of the function, e.g. the Gaussian distribution, the exponential distribution, etc. Most common distributions defined in the whole $\mathbb R^n$ domain have an exponential term that guarantees the PDF to integrate to unity by asymptotically reaching zero at infinity. In this case, it is possible to perform the calculations by working directly on the exponent of the distribution using logarithms. Without loss of generality, let us consider a Gaussian distribution $\mbf \phi_{\mbf x}(\mbf x,t_0)$, with mean $\boldsymbol \mu_0$ and covariance $\mbf P_0$:
\begin{equation}
 \phi_{\mbf x}(\mbf x,t_0) = \dfrac{\exp \Big(-\dfrac{1}{2} (\mbf x - \boldsymbol \mu_0)^T \mbf P_0^{-1} (\mbf x - \boldsymbol \mu_0) \Big)}{ (2\pi)^{n/2}\sqrt{\det \mbf P_0}}
\end{equation}
The exponent represents a separate function of the state that is obtained after applying the logarithm operator and disregarding the normalizing constant:
\begin{align}
     \zeta_{\mbf x}(\mbf x,t_0) &= -\dfrac{1}{2} (\mbf x - \boldsymbol \mu_0)^T \mbf P_0^{-1} (\mbf x - \boldsymbol \mu_0) \\
    &\overset{+}{=} \log (\phi_{\mbf x}(\mbf x,t_0))
\end{align}
where $\overset{+}{=}$ indicates an equality modulo an additive constant. The propagation of the Gaussian distribution using the KO approximation of the dynamics is performed as 
\begin{equation}
    \zeta_{\mbf x}(\mbf x,t_f) \overset{+}{=} \zeta_{\mbf x}(\mathcal W_{t_f\rightarrow t_0} (\mbf x),t_0) \label{eq:6}
\end{equation}
The \textit{posterior} distribution of the state (modulo a constant) is the evaluation of the \textit{prior} distribution of the state (modulo a constant) by the given inverted KOSTPM. 

The logarithmic reduction, whenever possible, greatly simplifies calculations as $\zeta_{\mbf x}(\mbf x,t_0)$ is a much easier function to evaluate. Conceptually, propagating the PDF according to Eq. \eqref{eq:5} or Eq. \eqref{eq:6} matters only if it is important to conserve the normalizing factor of the propagated PDF, as $\zeta_{\mbf x}(\mbf x,t_f)$ and $\psi_{\mbf x}(\mbf x,t_f)$ have the same shape. 

\section{Recursivity and Least Square Reduction} \label{sec5}
Whenever propagating distributions, being able to propagate the PDF iteratively while still accessing the distribution at each time step is fundamental. Especially in filtering, updating the PDF with the most current measurement is the core of estimation. Therefore, an additional algorithm must be followed to guarantee the tractability and recursivity of the proposed KO prediction method. 

The KOSTPM is a polynomial map with maximum order $\alpha$, which is arbitrarily selected. The inverted polynomial map has the same maximum order in its monomials. Each time a propagation is performed, the complexity of the function that represents the PDF increases drastically, especially due to the curse of dimensionality. For example, consider the Gaussian scenario. Using the logarithmic simplification, function $\zeta_{\mbf x}(\mbf x,t_0)$ is a second-order polynomial, as described by the quadratic formulation. When undergoing integration according to Eq. \eqref{eq:6}, the result is the propagated distribution $\zeta_{\mbf x}(\mbf x,t_f)$, which is a polynomial of order $2\alpha$. The evaluation of the function with the polynomial substitution doubled the maximum order of the approximation. It is evident that, to keep the procedure tractable and avoid the continuous order increase, a function complexity reduction must be implemented. 

Therefore, a polynomial reduction to represent the PDF is here derived, following a nonintrusive least square approach typical of polynomial chaos expansions \cite{ls}. First, $N$ independent and identically distributed samples are drawn uniformly from the PDF support: $\mbf x^{(j)} \quad \forall \quad j = 1,\dots,N$. Then, the probability of each point is evaluated from the predicted distribution
\begin{equation}
    p^{(j)} = \zeta_{\mbf x}(\mbf x^{(j)},t_f) \quad \forall \quad j = 1,\dots,N
\end{equation}
and stacked together as separate components to create 
\begin{equation}
    \mbf p = 
    \begin{bmatrix}
        p^{(1)} & p^{(2)} & \dots & p^{(N)}
    \end{bmatrix}^T
\end{equation}
which is the vector of realizations describing the probability of each sample. The points can now be used to fit a polynomial of a given order, $\omega < 2\alpha$, to match their probability. 
\begin{align}
    p^{(i)} &= \tilde \zeta_{\mbf x}(\mbf x,t_f) \nonumber \\
    &= c_0 + \sum_{r=1}^{\omega}\sum_k c_k x_1^{\gamma_1}\dots  x_n^{\gamma_r} \label{eq:7}
\end{align}
where the first summation indicates every order of the monomials up to $\omega $, while the second summation indicates every possible permutation of the exponents $\gamma \in \{1,\dots,n\}$. Conceptually, we are estimating the vector of coefficients $\mbf c$ that dictates the importance of each possible monomial in the representation of the propagated PDF. Equation \eqref{eq:7} is linear in the coefficients and can be written as \cite{least}
\begin{equation}
    \mbf p = \boldsymbol \Xi \mbf c
\end{equation}
where 
\begin{equation}
    \boldsymbol \Xi = \begin{bmatrix}
        1 & \mbf x_1^{(1)} & \dots & \mbf x_n^{(1)} & (\mbf x_n^{(1)})^2 & \mbf x_1^{(1)}\mbf x_2^{(1)} & \dots \\
        1 & \mbf x_1^{(2)} & \dots & \mbf x_n^{(2)} & (\mbf x_n^{(2)})^2 & \mbf x_1^{(2)}\mbf x_2^{(2)} & \dots \\
        \vdots & \vdots & \dots & \vdots & \vdots & \vdots &  \dots & \\
        1 & \mbf x_1^{(N)} & \dots & \mbf x_n^{(N)} & (\mbf x_n^{(N)})^2 & \mbf x_1^{(N)}\mbf x_2^{(N)} & \dots \\
    \end{bmatrix}
\end{equation}
Therefore, the set of coefficients $\mbf c $ for the polynomial approximation is estimated through least squares. Consider the following least square cost function 
\begin{equation}
    \mathcal J (\mbf c) = \left(\mbf p - \boldsymbol \Xi \mbf c \right)^T\left(\mbf p - \boldsymbol \Xi \mbf c \right)
\end{equation}
the optimal estimate is obtained by setting the condition
\begin{equation}
    \dfrac{\partial \mathcal J (\mbf c) }{\partial \mbf c} = \mbf 0
\end{equation}
which leads to the solution
\begin{equation}
    \mbf c = \left( \boldsymbol \Xi^T \boldsymbol \Xi\right)^{-1} \boldsymbol\Xi^T \mbf p
\end{equation}
Thanks to the coefficients, the propagated distribution $\zeta_{\mbf x}(\mbf x,t_f) $ is reduced to $\tilde \zeta_{\mbf x}(\mbf x,t_f) $ and a new prediction using the inverted KOSTPM map can be performed.

\section{Numerical Example} \label{sec6}
The proposed technique is tested in a numerical application: the propagation of a Gaussian PDF under the nonlinear dynamics of the Duffing oscillator. The Duffing oscillator is a system that describes the motion of a mass $m$ attached to a nonlinear spring, with a given stiffness coefficient $\kappa$, and a damper. The resulting equations of motion can be written in the form:
\begin{align}
    \dfrac{d}{d t} \mbf x = \begin{bmatrix}
        \dot x_1 \\ \dot x_2
    \end{bmatrix}= 
    \begin{bmatrix}
        \dfrac{x_2}{m} \\ -\kappa x_1 - \kappa a^2 \epsilon x_1^3
    \end{bmatrix}
\end{align}
where $a$ is a unit transformation constant and $\epsilon$ is a small parameter (chosen as 0.01), which introduces nonlinearity in the system. The initial PDF of the state is assumed Gaussian with mean $\hat{\mbf x}_0 = \begin{bmatrix}  0.4 & 0.6 \end{bmatrix}^T$ and covariance matrix $\mbf P_0 = \sigma_{\mbf x\mbf x}^2 \mbf I_{2\times2}$ with $\sigma_{\mbf x\mbf x} = 0.1$ and where $\mbf I_{2\times2}$ is the identity matrix. The distribution is propagated for $t_f = 500$ seconds, such that the resulting distribution is heavily non-Gaussian. 

\begin{figure}[!htb]
    \centering
    \includegraphics[width=0.9\linewidth]{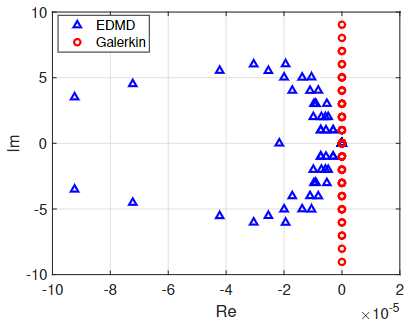}
    \caption{Eigenvalue comparison between EDMD and Galerkin.}
    \label{fig:eig}
\end{figure}
The results displayed in this application have been calculated following both the numerical and the analytical derivation of the KOSTPM. The two approaches give the same results as the evaluation of the Koopman matrices, and the projections from the inner products stand close. The eigenvalues of the system are reported in Fig. \ref{fig:eig}, where the Koopman matrix of order nine has been diagonalized. The Duffing oscillator is a pure periodic system, meaning that the true eigenvalues have solely imaginary parts. The Galerkin approach, reported as red circles, perfectly describes the eigendecomposition of the system as the eigenvalues populate the imaginary axis. As the Koopman order increases, more eigenvalues on the axis are obtained farther from the origin. Indeed, it can be noted that the KO order nine has nine eigenvalues in the positive and negative direction of the imaginary axis. This result comes from the eigenvalues composition of the original linear solution of the system, which can be obtained analytically. The linear eigenvalues are $[0, i]$ and $[0, -i]$, and the higher-order eigenvalues are obtained as every possible combination of the eigenvalues of the previous order. 
\begin{figure}[!htb]
    \centering
    \includegraphics[width=0.9\linewidth]{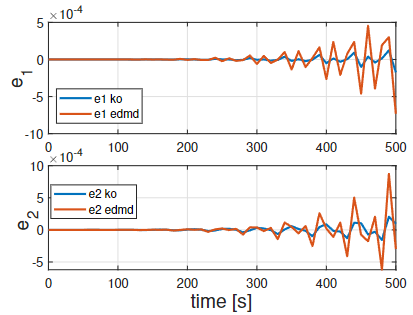}
    \caption{Solution accuracy comparison between EDMD and KO.}
    \label{fig:sol}
\end{figure}
Figure \ref{fig:eig} reports the EDMD eigenvalues as blue triangles. Their location is less precise when compared to the Galerkin counterparts and they have a non-zero real component, even if extremely small. This difference in the eigenvalue accuracy has consequences in the evaluation of the solution of the system. Figure \ref{fig:sol} shows the error $\mbf e$ of the KO propagation of the state of the system when compared to a high-accuracy Runge-Kutta integrator. The error becomes larger as the propagation time increases since the KOSTPM is less accurate. The figure shows that the Galerkin accuracy error, in blue, increases at a slower rate than the EDMD error, in red.

\begin{figure}[!htb]
    \centering
    \includegraphics[width=1.0\linewidth]{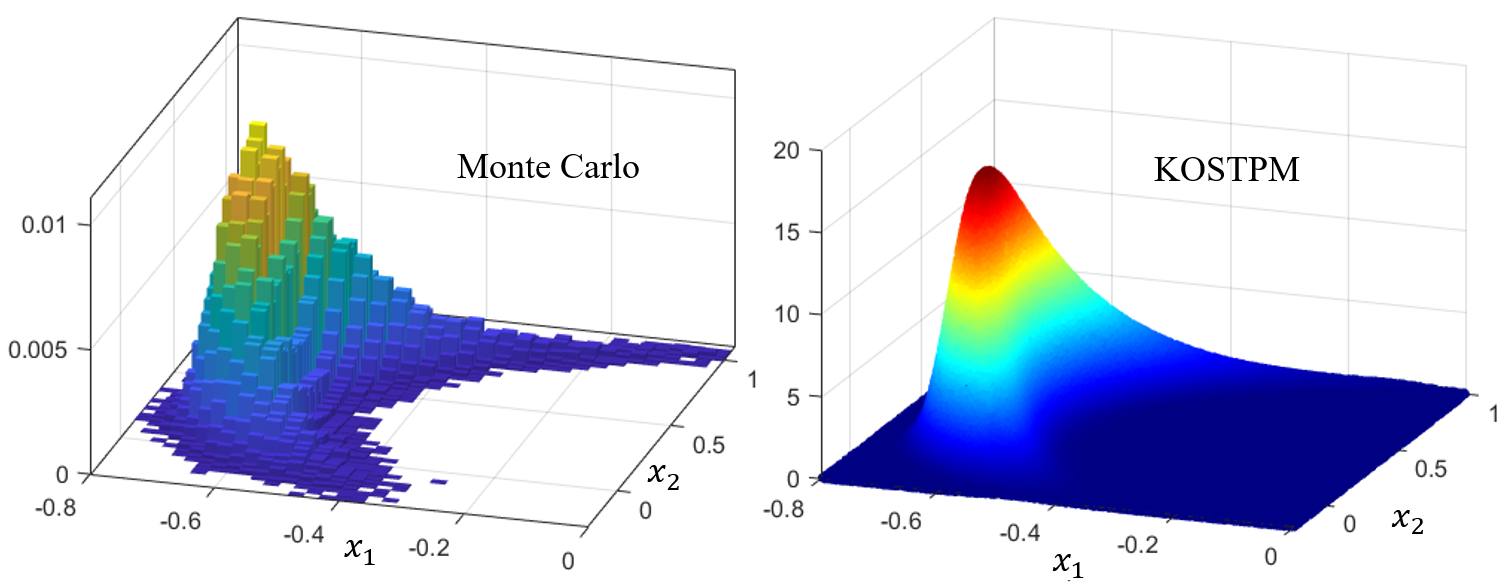}
    \caption{PDF KOSTPM propagation with Monte Carlo validation}
    \label{fig:3d}
\end{figure}
The main scope of the paper is the propagation of uncertainty. The prior Gaussian distribution is propagated to $t_f$ using the inverted KOSTPM and evaluated in the state space domain. The result, displayed in Fig. \ref{fig:3d}, shows the three-dimensional approximation of the propagated PDF. After undergoing a nonlinear system for a long propagation time, the distribution becomes non-Gaussian and bends following the dynamics. The KO approximation is validated by performing a Monte Carlo analysis, where one million samples from the prior distribution are propagated. Their final distribution is shown on the left side of Fig. \ref{fig:3d}. 
\begin{figure*} [!htb]
    \centering
    \includegraphics[width=0.92\textwidth]{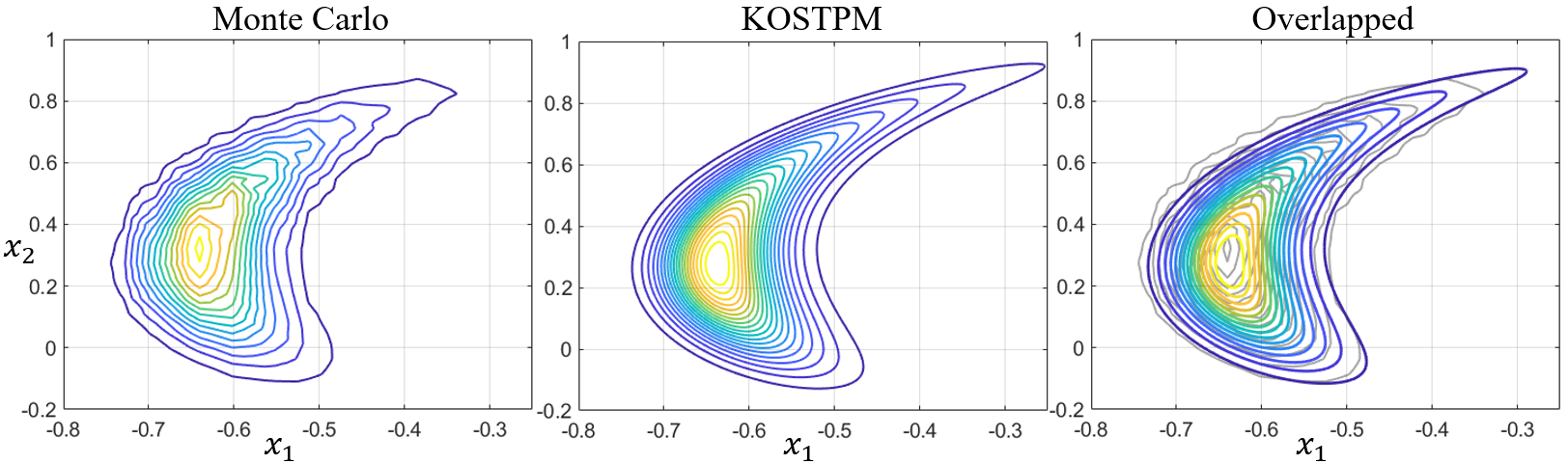}
    \caption{PDF propagation accuracy comparison at $t_f$: KOSTPM vs Monte Carlo.}
    \label{fig:pdfs}
\end{figure*}
The two distributions have been compared by considering their contour lines in Fig. \ref{fig:pdfs}. The first graph (left) shows the shape of the propagated PDF from the Monte Carlo samples; the middle graph is the KOSTPM solution of the distribution; and, lastly, the graph on the right overlaps the two PDFs. This visualization assesses the validity of the proposed method as the predicted and true distributions overlap. Therefore, the KOSTPM provides an analytical approximation of the state PDF at the desired time step in the form of a linear composition of polynomials. 

\begin{figure*}
    \centering
    \includegraphics[width=0.92\textwidth]{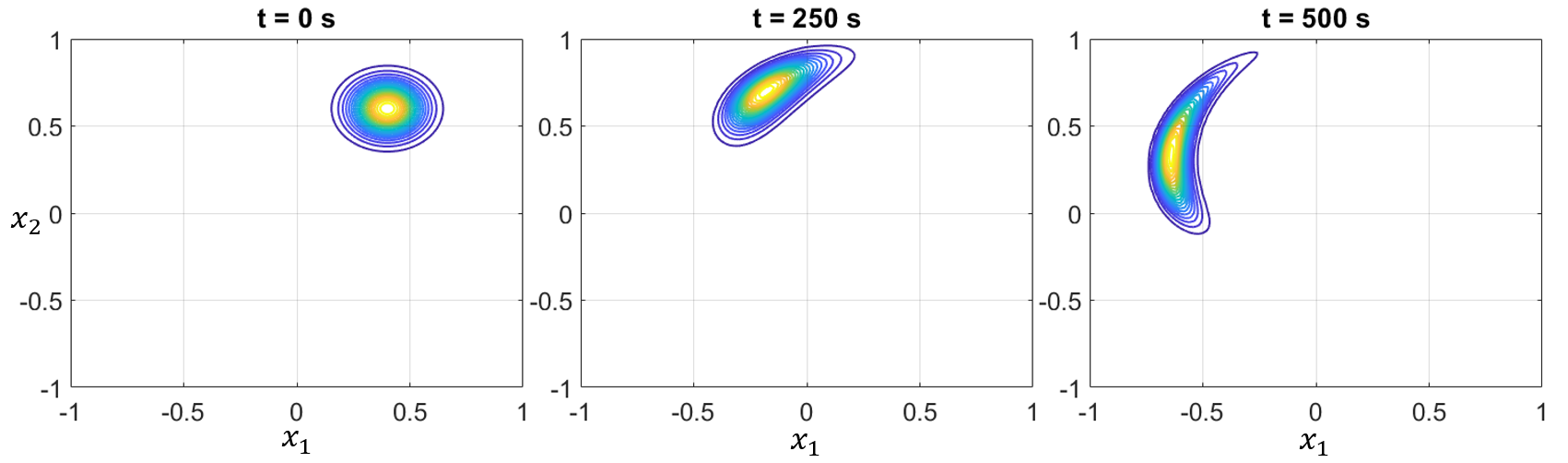}
    \caption{PDF KOSTPM propagation for multiple time steps.}
    \label{fig:time}
\end{figure*}
The subsequent propagation of PDFs is reported in Fig. \ref{fig:time},  which addresses the recursivity feature of the KO method. Starting from the initial Gaussian condition of the previous example (left contour), a first propagation of 250 seconds is performed (middle contour). Considering the logarithmic simplification and a 9th-order KOSTPM solution, the KO approximation of the PDF is an 18th-order polynomial. This representation is reduced to a 4th-order function via the least square reduction, such that a second propagation of 250 seconds can be carried out. Thus, the right contour of Fig. \ref{fig:time} reports the final state PDF, evaluated with a double propagation of half the time step. Assessing recursivity is critical in the uncertainty quantification problem, as having access to the predicted distribution allows its modification, which is fundamental to the Kalman filtering problem.

\section{Conclusion} \label{concl}
The Koopman approximation of the solution flow of the dynamics of the system has been proven helpful in propagating uncertainties. Thanks to the eigenfunction decomposition of the system, where the dynamics are projected on a set of well-defined basis functions, the inversion of the solution map is carried out by merely going backward in time. This solution leads to the evaluation of the propagated PDF as a simple polynomial evaluation. The numerical application showed the validity of the proposed approach, regardless of the selected methodology to evaluate the Koopman variables, either numerical via EDMD or analytical via Galerkin inner products. 

Recursivity of the process has been achieved by developing a reduction method based on the least square mathematics to keep the algorithm tractable. Access to the state PDF at each time step opens the technique to implementing the measurement update, which will be addressed in future developments. Therefore, the technique derived in this paper proposed a filtering prediction step that propagates directly the PDF rather than its first two central moments. 

\bibliographystyle{ieeetr}
\bibliography{references.bib}


\end{document}